\documentclass[%
reprint,
superscriptaddress,
amsmath,
amssymb,
aps,
]{revtex4-1}

\usepackage{graphicx}
\usepackage{dcolumn}
\usepackage{bm}
\usepackage[export]{adjustbox}
\usepackage{xcolor,soul}

\begin{document}

\preprint{APS/123-QED}

\title{The MICROSCOPE mission: first results of a space test of the Equivalence Principle}

\author{Pierre Touboul}
\email{Pierre.Touboul@onera.fr}
\affiliation{ONERA, chemin de la Huni\`ere, BP 80100, F-91123 Palaiseau Cedex, France}
\author{Gilles M\'etris}%
\email{Gilles.Metris@oca.eu }
\affiliation{Universit\'e C\^ote d{'}Azur, Observatoire de la C\^ote d'Azur, CNRS, IRD, G\'eoazur, 250 avenue Albert Einstein, F-06560 Valbonne, France}
\author{Manuel Rodrigues}%
\email{Manuel.Rodrigues@onera.fr }
\affiliation{ONERA, chemin de la Huni\`ere, BP 80100, F-91123 Palaiseau Cedex, France}

\author{Yves Andr\'e}%
\affiliation{CNES, 18 avenue Edouard Belin, F-31401 Toulouse, France}

\author{Quentin Baghi}%
\affiliation{Universit\'e C\^ote d{'}Azur, Observatoire de la C\^ote d'Azur, CNRS, IRD, G\'eoazur, 250 avenue Albert Einstein, F-06560 Valbonne, France}

\author{Jo\"el Berg\'e}%
\affiliation{ONERA, chemin de la Huni\`ere, BP 80100, F-91123 Palaiseau Cedex, France}
\author{Damien Boulanger}%
\affiliation{ONERA, chemin de la Huni\`ere, BP 80100, F-91123 Palaiseau Cedex, France}
\author{Stefanie Bremer}%
\affiliation{ZARM, Center of Applied Space Technology and Microgravity, University of Bremen, Am Fallturm, D-28359 Bremen, Germany}
\author{Patrice Carle}%
\affiliation{ONERA, chemin de la Huni\`ere, BP 80100, F-91123 Palaiseau Cedex, France}
\author{Ratana Chhun}%
\affiliation{ONERA, chemin de la Huni\`ere, BP 80100, F-91123 Palaiseau Cedex, France}
\author{Bruno Christophe}%
\affiliation{ONERA, chemin de la Huni\`ere, BP 80100, F-91123 Palaiseau Cedex, France}
\author{Valerio Cipolla}%
\affiliation{CNES, 18 avenue Edouard Belin, F-31401 Toulouse, France}
\author{Thibault Damour}%
\affiliation{IHES, Institut des Hautes \'Etudes Scientifiques, 35 route de Chartres, F-91440 Bures-sur-Yvette, France}
\author{Pascale Danto}%
\affiliation{CNES, 18 avenue Edouard Belin, F-31401 Toulouse, France}
\author{Hansjoerg Dittus}%
\affiliation{DLR, K\"oln headquarters, Linder H\"ohe, 51147 K\"oln, Germany}
\author{Pierre Fayet}%
\affiliation {LPTENS, \'Ecole Normale Sup\'erieure (PSL Research University), 24 rue Lhomond, 75231 Paris Cedex 05, France}
\author{Bernard Foulon}%
\affiliation{ONERA, chemin de la Huni\`ere, BP 80100, F-91123 Palaiseau Cedex, France}
\author{Claude Gageant}%
\affiliation{ONERA, chemin de la Huni\`ere, BP 80100, F-91123 Palaiseau Cedex, France}
\author{Pierre-Yves Guidotti}%
\affiliation{CNES, 18 avenue Edouard Belin, F-31401 Toulouse, France}
\author{Daniel Hagedorn}%
\affiliation {PTB, Physikalisch-Technische Bundesanstalt, Bundesallee 100, 38116 Braunschweig, Germany}
\author{Emilie Hardy}%
\affiliation{ONERA, chemin de la Huni\`ere, BP 80100, F-91123 Palaiseau Cedex, France}
\author{Phuong-Anh Huynh}%
\affiliation{ONERA, chemin de la Huni\`ere, BP 80100, F-91123 Palaiseau Cedex, France}
\author{Henri Inchauspe}%
\affiliation{ONERA, chemin de la Huni\`ere, BP 80100, F-91123 Palaiseau Cedex, France}
\author{Patrick Kayser}%
\affiliation{ONERA, chemin de la Huni\`ere, BP 80100, F-91123 Palaiseau Cedex, France}
\author{St\'ephanie Lala}%
\affiliation{ONERA, chemin de la Huni\`ere, BP 80100, F-91123 Palaiseau Cedex, France}
\author{Claus L\"ammerzahl}%
\affiliation {ZARM, Center of Applied Space Technology and Microgravity, University of Bremen, Am Fallturm, D-28359 Bremen, Germany}
\author{Vincent Lebat}%
\affiliation{ONERA, chemin de la Huni\`ere, BP 80100, F-91123 Palaiseau Cedex, France}
\author{Pierre Leseur}%
\affiliation{ONERA, chemin de la Huni\`ere, BP 80100, F-91123 Palaiseau Cedex, France}
\author{Fran\c coise Liorzou}%
\affiliation{ONERA, chemin de la Huni\`ere, BP 80100, F-91123 Palaiseau Cedex, France}
\author{Meike List}%
\affiliation {ZARM, Center of Applied Space Technology and Microgravity, University of Bremen, Am Fallturm, D-28359 Bremen, Germany}
\author{Frank L\"offler}%
\affiliation {PTB, Physikalisch-Technische Bundesanstalt, Bundesallee 100, 38116 Braunschweig, Germany}
\author{Isabelle Panet}%
\affiliation{IGN, Institut g\'eographique national, 73 avenue de Paris, F-94160 Saint Mand\'e, France}
\author{Benjamin Pouilloux}%
\affiliation{CNES, 18 avenue Edouard Belin, F-31401 Toulouse, France}
\author{Pascal Prieur}%
\affiliation{CNES, 18 avenue Edouard Belin, F-31401 Toulouse, France}
\author{Alexandre Rebray}%
\affiliation{ONERA, chemin de la Huni\`ere, BP 80100, F-91123 Palaiseau Cedex, France}
\author{Serge Reynaud}%
\affiliation{Laboratoire Kastler Brossel, UPMC-Sorbonne Universit\'e, CNRS, ENS-PSL Research University, Coll\`ege de France, F-75005 Paris, France}
\author{Benny Rievers}%
\affiliation{ZARM, Center of Applied Space Technology and Microgravity, University of Bremen, Am Fallturm, D-28359 Bremen, Germany}
\author{Alain Robert}%
\affiliation{CNES, 18 avenue Edouard Belin, F-31401 Toulouse, France}
\author{Hanns Selig}%
\affiliation{ZARM, Center of Applied Space Technology and Microgravity, University of Bremen, Am Fallturm, D-28359 Bremen, Germany}
\author{Laura Serron}%
\affiliation{Universit\'e C\^ote d{'}Azur, Observatoire de la C\^ote d'Azur, CNRS, IRD, G\'eoazur, 250 avenue Albert Einstein, F-06560 Valbonne, France}
\author{Timothy Sumner}%
\affiliation{Blackett Laboratory, Imperial College London, United Kingdom}
\author{Nicolas Tanguy}%
\affiliation{ONERA, chemin de la Huni\`ere, BP 80100, F-91123 Palaiseau Cedex, France}
\author{Pieter Visser}%
\affiliation{Faculty of Aerospace Engineering, Delft University of Technology, Kluyverweg 1, 2629 HS Delft, The Netherlands}

\date{\today}

\begin{abstract}
According to the Weak Equivalence Principle, all bodies should fall at the same rate in a gravitational field. The MICROSCOPE satellite, launched in April 2016, aims to test its validity at the $10^{-15}$ precision level, by measuring the force required to maintain two test masses (of titanium and platinum alloys) exactly in the same orbit. A non-vanishing result would correspond to a violation of the Equivalence Principle, or to the discovery of a new long-range force. Analysis of the first data gives  $\delta\rm{(Ti,Pt)}= [-1 \pm 9 (\mathrm{stat}) \pm 9 (\mathrm{syst})] \times 10^{-15}$ (1$\sigma$ statistical uncertainty) for the titanium-platinum E\"otv\"os parameter characterizing the relative difference in their free-fall accelerations. 

\end{abstract}

\maketitle


\section{\label{sec:level1}Introduction}

Gravity seems to enjoy a remarkable universality property: bodies of different compositions fall at the same rate in an external gravitational field~\cite{eotvos22, wagner12, williams12}. Einstein interpreted this as an equivalence between gravitation and inertia~\cite{einstein07}, and used this (Weak) Equivalence Principle (WEP) as the starting point for the theory of General Relativity~\cite{einstein16}. In terms of the E\"otv\"os parameter $\delta\rm{(A,B)}=2(a_A-a_B)/(a_A+a_B)$ ($a_A$ and $a_B$ being the free-fall accelerations of the two bodies A and B), the best laboratory ($1\sigma$) upper limits on $\delta\rm{(A,B)}$ are $\delta\rm{(Be,Ti)} = (0.3 \pm 1.8) \times 10^{-13}$ and  
$\delta\rm{(Be,Al)} = (-0.7 \pm 1.3) \times 10^{-13}$~\cite{wagner12}, with similar limits on the differential acceleration between the Earth and the Moon toward the Sun~\cite{williams12}.

General Relativity (GR) has passed all historical and current experimental tests~\cite{will14}, including, most recently, the direct observation of the gravitational waves emitted by two coalescing black holes~\cite{abbott16}. However, it does not provide a consistent quantum gravity landscape and leaves many questions unanswered, in particular about dark energy and the unification of all fundamental interactions. 
Possible avenues to close those problems may involve very weakly coupled new particles, such as the string-theory spin-0 dilaton~\cite{damour94,damour02}, a chameleon~\cite{khoury04} or a spin-1 boson U from an extended gauge group~\cite{fayet90,fayet17}, generally leading to an apparent WEP violation. 

The MICROSCOPE space mission implements a new approach to test the WEP by taking advantage of the very quiet space environment.
Non-gravitational forces acting on the satellite are counteracted by cold gas thrusters making it possible to compare the accelerations of two test masses of different compositions ``freely-falling" in the same orbit around the Earth for a long period of time~\cite{touboul01,touboul12}. This is done by accurately measuring the force required to keep the two test masses in relative equilibrium. Present data allow us to improve the $1\sigma$ upper limit on the validity of the WEP by an order of magnitude.

\section{\label{sec2}The MICROSCOPE space mission}

MICROSCOPE aims to test the Equivalence Principle with an unprecedented precision of $10^{-15}$. The T-SAGE (Twin Space Accelerometers for Gravitation Experiment) scientific payload, provided by ONERA, is integrated within a CNES microsatellite. It was launched and injected into a 710\,km altitude, circular orbit, by a Soyouz launcher from Kourou on April 25, 2016. The orbit is sun-synchronous, dawn-dusk (i.e. the ascending node stays at 18\,h mean solar time) in order to have long eclipse-free periods (eclipses are defined as periods within the Earth{'}s shadow and happen only between May and July).

T-SAGE is composed of two parallel similar differential accelerometer instruments, each one with two concentric hollow cylindrical test-masses. They are exactly the same, except for the use of different materials for the test-masses. In one instrument (SUREF) the two test-masses have the same composition, and are made from a Platinum/Rhodium alloy (90/10). In the other instrument (SUEP) the test-masses have different compositions: Pt/Rh~(90/10) for the inner test-mass and Titanium/Aluminum/Vanadium~(90/6/4) (TA6V) for the outer test-mass (see Table~\ref{tab:masses}). The test-masses' shape has been designed to reduce the local self-gravity gradients due to multipole moment residues~\cite{connes97,willemenot97}.
\begin{table}[b]
\caption{\label{tab:masses}%
Main test-mass physical properties measured in the laboratory before integration in the instrument.
}
\begin{ruledtabular}
\begin{tabular}{lcccc}
\textrm{Measured}&
\textrm{SUREF}&
\textrm{SUREF}&
\textrm{SUEP} & SUEP\\
parameters & Inner mass & Outer mass & Inner mass & Outer mass \\
at $20\,^o$C & Pt/Rh & Pt/Rh & Pt/Rh & Ti/Al \\
\hline

Mass in kg & 0.401533 & 1.359813 & 0.401706 & 0.300939\\ \hline
Density in  & 19.967 & 19.980 & 19.972 & 4.420\\
g\,cm$^{-3}$ &  &  &  & \\
\end{tabular}
\end{ruledtabular}
\end{table}

The test-masses experience almost the same Earth gravity field and are constrained by electrostatic forces to follow the same quasi-circular orbit. A WEP violation ($\delta\rm{(A,B)} \neq 0$) would result in a difference $-\delta\rm{(A,B)}\overrightarrow{g}$ in the electrostatic feedback forces providing the accelerations needed to maintain the test masses in the same orbit. The satellite can be spun around the normal to the orbital plane and oppositely to the orbital motion in order to increase the frequency of the Earth gravity modulation. 
In this case, in the satellite frame, the Earth gravity field rotates at the sum of the orbital and spin frequencies (see Fig. 1). 
A WEP violation would give a signal modulated at this frequency, denoted $f_{\mathrm{EP}}$. The Earth gravity field has a mean amplitude of 7.9\,m\,s$^{-2}$ at 710\,km altitude, and testing the WEP with an accuracy of $10^{-15}$ necessitates measuring the differential constraining force per unit of mass (henceforth called acceleration) between test mass pairs with an $1\sigma$ accuracy of $7.9\times 10^{-15}$\,m\,s$^{-2}$ at $f_{\mathrm{EP}}$. 

SUEP and SUREF use servo-loops to maintain each test mass motionless with respect to its surrounding silica electrodes, with a relative position resolution of  $3\times 10^{-11}$\,m\,Hz$^{-1/2}$ measured within the bandwidth [$2\times 10^{-4}$\,Hz, 1\,Hz]. 
The position measurement noise leads to an acceleration noise contribution lower than $2\times 10^{-14}$\,m\,s$^{-2}\,$Hz$^{-1/2}$ at frequencies $f_{\mathrm{EP}}=3.1113\times10^{-3}$\,Hz (for SUEP) and $f_{\mathrm{EP}}=0.9250\times 10^{-3}$\,Hz (for SUREF).  This is well below the requirement specification of $2\times 10^{-12}$\,m\,s$^{-2}$\,Hz$^{-1/2}$ at $f_{\mathrm{EP}}$ for each instrument. The electrode sets are engraved on silica parts whose positions are very stable with respect to a common silica "hat" part mounted on a common INVAR sole plate. Electrostatic forces are exerted capacitively on the test masses without any mechanical contact.  Thin gold wires of $7\,\mu$m diameter are used to control the charge on each test-mass.

\begin{figure*}
\includegraphics[width=0.45 \textwidth, trim=0cm 0cm 0cm 0cm, clip=true, frame]{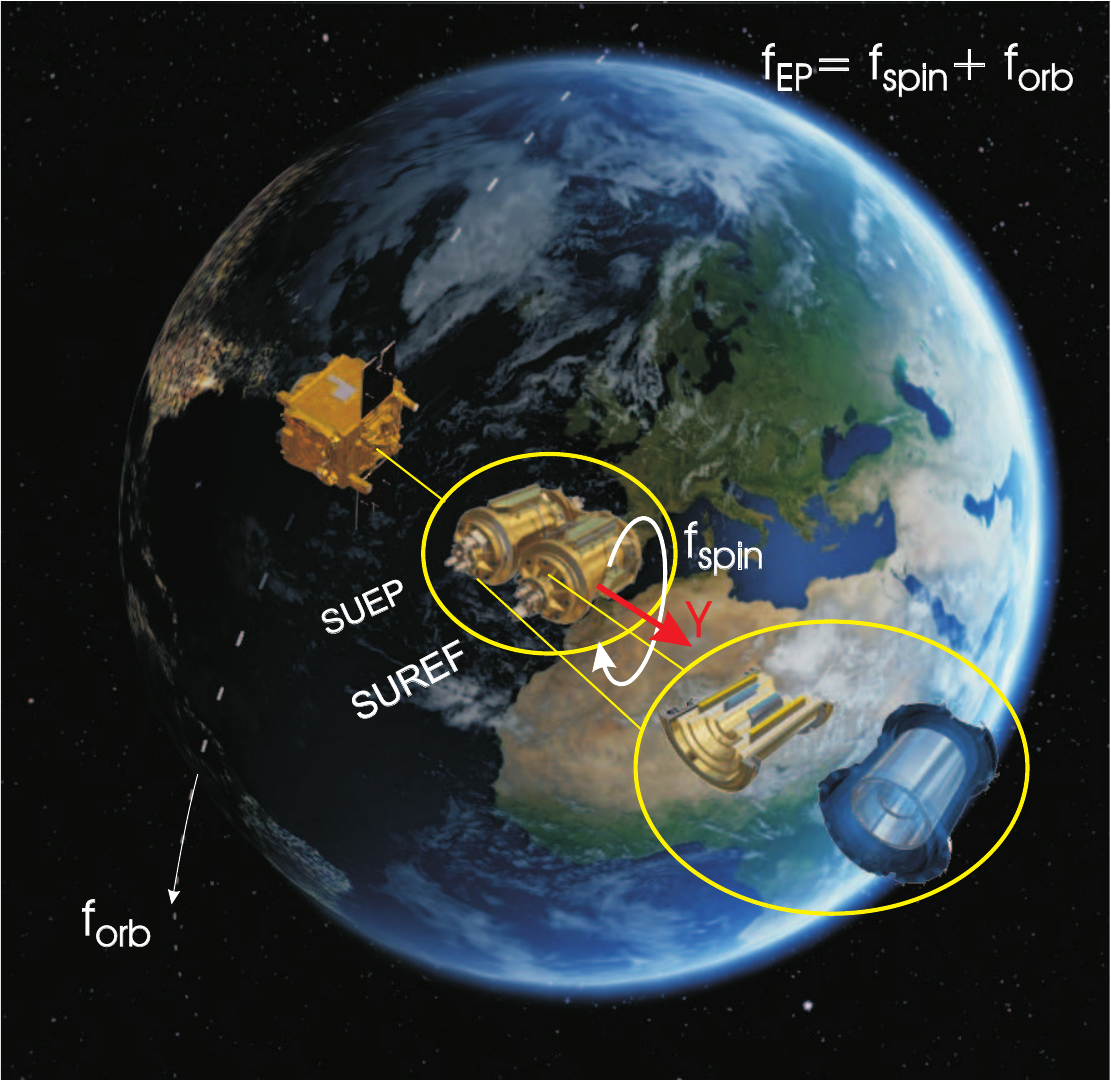}
\includegraphics[width=0.453 \textwidth, trim=-1cm -0.5cm -1cm -0.5cm, clip=true, frame]{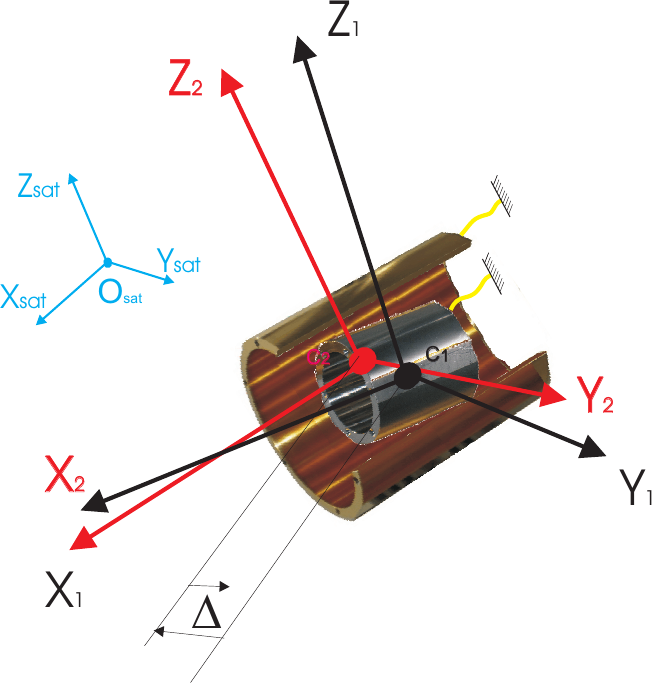}
\caption{\label{fig:orbit} Left: the  4 test-masses orbiting around the Earth (credits CNES / Virtual-IT 2017). Right: test-masses and satellite frames; the ($X_{\rm sat}$, $Y_{\rm sat}$, $Z_{\rm sat}$) triad defines the satellite frame; the reference frames ($X_k$, $Y_k$, $Z_k$, $k=1,2$) are attached to the test-masses (black for the inner mass $k=1$, red for the outer mass $k=2$); the $X_k$ axes are the test-mass cylinders' longitudinal axis and define the direction of WEP test measurement; the $Y_k$ axes are normal to the orbital plane, and define the rotation axis when the satellite spins; the $Z_k$ axes complete the triads.  The 7 $\mu$m gold wires connecting the test-masses to the common Invar sole plate are shown as yellow lines. $\vec{\Delta}$ represents the test-masses offcentering. The centers of mass corresponds to the origins of the sensor-cage-attached reference frames.}
\end{figure*}

Both high-frequency (100 kHz~\cite{josselin99}) capacitive sensing and low-frequency ($<1$\,Hz) control of each test-mass' position and attitude about its 6 degrees of freedom (DoF) are performed by the same set of electrodes. The position and attitude are derived from the combination of different electrodes' capacitive sensing, then a digital PID (Proportional Integral Derivative) control calculates the necessary voltage to apply to each electrode. For each pair of symmetric electrodes controlling one DoF, the (small) antisymetric voltages applied on the electrodes are superimposed on a larger DC voltage, thereby making the applied electrostatic forces proportional to first order to the applied voltages. The output of the instrument is thus derived from the applied voltages.  
In the absence of a WEP violation, and if everything is perfect and aligned (in contrast to the exagerated case of Fig.~\ref{fig:orbit}'s right panel), the difference of accelerations of two concentric test-masses is expected to vanish whatever their composition or mass. In case of a violation, the difference of accelerations would be directly proportional to the magnitude of the Earth's gravitational field.

To improve the measurements, additional servo-loops reduce non-gravitational accelerations of the satellite for the six DoF using cold gas thrusters driven by the accelerometers' measurements of the linear and angular accelerations (similar to LISA Pathfinder~\cite{armano16}). The payload measurements are completed by satellite attitude measurements from the star trackers. The thrusters can also apply additional accelerations to the satellite in order to calibrate the instruments. 

During most of the scientific sessions the drag-free loop is controlled by the output of one of the test-masses. We have checked that the residual acceleration measurements were below $1.5\times 10^{-12}$\,m\,s$^{-2}\,$Hz$^{-1/2}$  for this test-mass and below $3\times 10^{-11}$\,m\,s$^{-2}\,$Hz$^{-1/2}$ for the other; this is much better than the requirement on the Drag-Free and Attitude Control System (DFACS) of $3\times 10^{-10}$\,m\,s$^{-2}\,$Hz$^{-1/2}$ about $f_{\mathrm{EP}}$. The DFACS limits the acceleration of one of the two instruments (SUEP or SUREF, depending of the session). The other instrument, 17.5\,cm away (mainly along the Y axis), undergoes inertial and gravity gradient accelerations which preclude getting the same performance despite the excellent attitude control of the satellite. This is one of the reasons why we conduct independent experiments in different sessions, using either SUREF or SUEP, but not both simultaneously.

The payload is integrated inside a magnetic shield at the center of the microsatellite whose efficiency was modeled with a 3D magnetic tool and with measured magnetic properties on instrument parts. The sensor geometry and the low noise electronics benefit from the very stable passive thermal cocoon of the satellite. 

\section{\label{sec3}Measurements and estimation of systematic errors}

We define ${\vec \Gamma_k} $ as the acceleration exerted by the surrounding capacitive sensor cage on the $k$-th test-mass. The three components of each acceleration  ${\vec \Gamma_k} $ are measured in the frame ($X_k$,~$Y_k$,~$Z_k$) attached to the corresponding sensor cage (see Fig.~\ref{fig:orbit}). Because of small (time-independent) misalignments with respect to the satellite frame  ($X_{\rm sat}$,~$Y_{\rm sat}$,~$Z_{\rm sat}$), the locally measured components ${\vec \Gamma_k} $ are related to their components ${\vec \Gamma_k^{\rm sat}} $ in the satellite frame via
${\vec \Gamma_k}  = [\theta_k] {\vec \Gamma_k^{\rm sat}} $,  where the matrix $[\theta_k]$ reads $\left[\theta_{k}\right] = \left[
\begin{matrix}
1 & \theta_{kz} & -\theta_{ky} \\
-\theta_{kz} & 1 & \theta_{kx} \\
\theta_{ky} & -\theta_{kx} & 1 \\
\end{matrix}
\right]  $.
The three (antisymmetric) off-diagonal elements $\theta_{kl}$ measure the small rotation between the satellite frame and the $k$-th test-mass frame (designed such that $\theta_{kl} < 2.5 \times 10^{-3}$\,rad).

Besides the antisymmetric off-diagonal elements $\theta_{kl}$ there are also measurement biases, non-unit scale factors and coupling defects which lead the readouts to measure the components ${\vec \Gamma_k}^{\mathrm{meas}} = [A_k] {\vec \Gamma_k}$ where the sensitivity  matrix $[A_k]$ reads $\left[A_k \right]~=~\underbrace{\left[
\begin{matrix}
1+K_{kx} & 0 & 0 \\
0 & 1+K_{ky} & 0 \\
0 & 0 & 1+K_{kz}
\end{matrix}
\right]}_{\text{scale factor}} + \underbrace{\left[
\begin{matrix}
0 & \eta_{kz} & \eta_{ky} \\
\eta_{kz} & 0 & \eta_{kx} \\
\eta_{ky} & \eta_{kx} & 0
\end{matrix}
\right]}_{\text{coupling}}$.

We then define the common- and differential-mode sensitivity matrices of the two inertial sensors as:
$\left[M_c\right] = \frac{1}{2}\left(\left[A_1\right]\left[\theta_1\right]+\left[A_2\right]\left[\theta_2\right]\right)$ and
$\left[M_d\right] = \frac{1}{2}\left(\left[A_1\right]\left[\theta_1\right]-\left[A_2\right]\left[\theta_2\right]\right)$. 
By design, the elements of $\left[M_d\right]$ are smaller than $10^{-2}$ and known to $10^{-4}$ accuracy after in-orbit estimation. Similarly, $\left[M_c\right]$ is close to the identity matrix with a subpercent error. 
 
The quantity of interest is the difference between the accelerations exerted on the two test-masses of a given sensor unit, namely the inner mass ($k=1$) and the outer mass ($k=2$), ${\vec \Gamma_d}^{\rm meas} \equiv {\vec \Gamma_1}^{\rm meas} - {\vec \Gamma_2}^{\rm meas}$.
This measured differential  acceleration is directly related to the E\"otv\"os ratio $\delta(2,1)$ and to the various forces acting on the satellite (see Ref. [20] for a detailed derivation):
\begin{eqnarray}
\overrightarrow\Gamma_d^{\mathrm{meas}} & \simeq & \overrightarrow K_{0,d}\\  \nonumber
&& + \left[M_c\right]\Big( \left( \left[ T \right]- \left[In\right]  \right)\overrightarrow\Delta -
2\left[\Omega \right]\dot{\overrightarrow\Delta}-\ddot{\overrightarrow\Delta}  \\  \nonumber
&& +\delta\left(2,1\right) \overrightarrow g 
\left( O_{\mathrm{sat}} \right) \Big) \\  \nonumber
&& + 2\left[ M_d\right] \overrightarrow\Gamma_c^{\mathrm{app}} + 
\overrightarrow \Gamma_d^{\mathrm{quad}} + \left[ \mathrm{Coupl}_d\right] \dot{\overrightarrow \Omega}+ \overrightarrow \Gamma_d^{n}.
\end{eqnarray}

\begin{table*}
\caption{\label{tab:terms}%
Description of the terms in Eq.\,(1).
}
\begin{ruledtabular}
\begin{tabular}{ll}
{\bf Terms of Eq.\,(1)} & {\bf Description of the terms} \\
\hline
$\overrightarrow K_{0,d}$ & Vector of the difference of the inertial sensor measurement bias. \\ \hline
$\overrightarrow \Delta = \left(\Delta_x,\Delta_y,\Delta_z \right)^T$ & Vector (in the satellite frame) connecting the center of the inner mass to that of the outer mass. \\ \hline
$\dot{\overrightarrow \Delta}$ and $\ddot{\overrightarrow \Delta}$ & First and second time derivatives of $\overrightarrow \Delta$. They are nullified in the instrument{'}s bandwidth\\
{} & when the instrument's servo-controls maintain the masses motionless versus the satellite frame. \\ \hline
$\left[ \Omega \right]$ & Satellite's angular velocity matrix, ${\vec \Omega} \times {\vec r} = [\Omega] \overrightarrow r$ \\ \hline
$\left[ T \right]$ & Gravity gradient tensor in the satellite frame. \\ \hline
$\left[ In \right]$ & Matrix gradient of inertia defined in the satellite frame by $\left[ In \right] = \left[\dot \Omega\right] + \left[ \Omega \right]\left[ \Omega \right]$ . \\ \hline
$\overrightarrow g =\left(g_x, g_y, g_z \right)^T$ & Gravity acceleration vector in the satellite frame of 7.9\,m\,s$^{-2}$ in magnitude at the 710\,km altitude.\\  \hline
$\delta \left(2,1 \right)$ & E\"otv\"os parameter of the outer mass (2) with respect to the inner mass (1).\\ \hline
$2\left[ \Omega \right]\dot{\overrightarrow \Delta}$  & Coriolis effect in the satellite frame. Very weak because the relative velocity of the test-masses \\ 
{} & at the test frequency is limited by the integral term of the accelerometer{'} servo-loops  and because \\ 
{} & the angular velocity is well controlled by the satellite DFACS loops.  \\ \hline
$\overrightarrow \Gamma_c^{\mathrm{app}}$ & Mean acceleration applied on both masses in the satellite frame. Limited by the satellite DFACS. \\ \hline
$\overrightarrow \Gamma_d^{\mathrm{quad}}$ & Difference of the non-linear terms in the measurement, mainly the difference of the quadratic responses of \\
{} & the inertial sensors.\\ \hline
$\left[ \mathrm{Coupl}_d \right]$ & Matrix of the difference, between the two sensors, of the coupling from the angular \\ {} & acceleration $\dot{\overrightarrow \Omega}$  to the linear acceleration.  \\ \hline
$\overrightarrow \Gamma_d^{n}$  & Difference of the acceleration measurement noises of the two sensors (coming from thermal noise,  \\
{} & electronics noise, parasitic forces,...), comprising stochastic and systematic error sources. 
\end{tabular}
\end{ruledtabular}
\end{table*}

All terms in Eq.\,(1) are described in Table~\ref{tab:terms}. Eq.\,(1) shows that the measurement may be sensitive to the common acceleration of the platform applied to both sensors of each instrument. Hence the mission scenario includes  calibration sessions scheduled to match the sensitivities of the sensors, in order to estimate $\left[M_d\right]$ and to a posteriori correct its effect~\cite{hardy13}. 

The gravity acceleration $\overrightarrow g$ and the gravity gradient tensor $\left[T \right]$ projected into the satellite frame are computed from the ITSG-GRACE2014s Earth's gravity potential model ~\cite{mayer06}, by using the measured position and attitude of the satellite.
The distance between the two test-masses' centers of mass is estimated to $(\Delta_x, \Delta_y, \Delta_z)=(20.1, -8.0, -5.6)\pm(0.1,0.2,0.1)\, \mu$m. The $\Delta_x$ and $\Delta_z$ components are estimated from the gravity gradient signal at $2f_{\mathrm{EP}}$ (at $2 f_{\mathrm{EP}}$, systematic errors are smaller than $8\times 10^{-14}$\,m\,s$^{-2}$, much smaller than raquired for the above $0.1\mu$m accuracy). The corresponding acceleration is simultaneously computed and corrected from the measured differential acceleration. The $\Delta_y$ component, although contributing only marginally to the differential acceleration, is estimated through a dedicated session \cite{hardy13}. In the particular mode where the satellite is spinning, the effect of test-mass miscentering is negligible at $f_{\mathrm{EP}}$ and could be left uncorrected.
The satellite orbit and attitude are determined to 0.42\,m and $0.4\,\mu$rad precision, much better than the required 2\,m and $1\,\mu$rad.

The different error source contributions to Eq.\,(1) are summarized in Table~\ref{tab:calibration} \cite{touboul09,touboul17}. As X is the preferred axis for the EP test, in-flight calibration of the first-row coefficients of $\left[M_d\right]$ is sufficient: $M_{dxx}$=$8.5\times 10^{-3}\pm 1.5\times 10^{-4}$, $|M_{dxy}|$ and $|M_{dxz}|<1.5\times 10^{-4}$\,rad.
The effect of the Earth{'}s gravity field and its gradient is considered along X at $f_{\mathrm{EP}}$ and in phase with any EP signal. All other terms are considered at $f_{\mathrm{EP}}$ but without considering the phase which is conservative. 

\begin{table*}
\caption{\label{tab:calibration}%
Evaluation of systematic errors in the differential acceleration measurement for SUEP @$f_{\mathrm{EP}}$=3.1113$\times10^{-3}\,Hz$.
}
\begin{ruledtabular}
\begin{tabular}{lll}
{\bf Term in the Eq.\,(1) projected on $\overrightarrow x$} & \vline {\bf  Amplitude or upper bound} & \vline {\bf Method of estimation} \\ \hline
{\bf Gravity gradient effect $\left[ T \right]\overrightarrow \Delta$ along X @ $f_{\mathrm{EP}}$} & {\bf (in phase with $g_x$)} & \\ \hline
($T_{xx}\Delta x$; $T_{xy}\Delta y$; $T_{xz}\Delta z$)  & \vline $< (10^{-18}$;$ 10^{-19}$;$ 10^{-17}$)\,m\,s$^{-2}$  & \vline Earth{'}s gravity model and in flight calibration. \\ \hline
{\bf Gradient of inertia matrix $\left[ In \right]$ effect along} & {\bf X @  $f_{\mathrm{EP}}$}  & \\ \hline
$\dot \Omega_y\Delta_z - \dot \Omega_z \Delta_y$ & \vline $10^{-18}$\,m\,s$^{-2}$ & \vline DFACS performances and calibration. \\ \hline
$\Omega_x \Omega_y \Delta_y - \Omega_x \Omega_z \Delta_z - \left(\Omega_y^2 + \Omega_z^2 \right)\Delta_x$  & 
\vline $1.3 \times 10^{-17}$\,m\,s$^{-2}$ & \vline DFACS performances and calibration. \\ \hline
{\bf Drag-free control} & {} & \\ \hline
$(\left[ M_d \right] \overrightarrow \Gamma_c^{\mathrm{app}}). \overrightarrow x$ & \vline $1.7 \times 10^{-15}$\,m\,s$^{-2}$ & \vline DFACS performances and calibration. \\ \hline
{\bf Instrument systematics and defects} & {} & \\ \hline
$(\overrightarrow \Gamma_d^{\mathrm{quad}}).\overrightarrow x$  & \vline $5 \times 10^{-17}$\,m\,s$^{-2}$ & \vline DFACS performances and calibration. \\ \hline
$([\mathrm{Coupl}_d] \dot {\overrightarrow \Omega}).\overrightarrow x$ & \vline $< 2 \times 10^{-15}$\,ms$^{-2}$ & \vline Couplings observed during commissioning phase.\\ \hline
Thermal systematics & \vline $< 67 \times 10^{-15}$\,m\,s$^{-2}$ & \vline Thermal sensitivity in-orbit evaluation. \\ \hline
Magnetic systematics & \vline $< 2.5 \times 10^{-16}$\,m\,s$^{-2}$ & \vline Finite elements calculation.\\ \hline
{\bf Total of systematics in $\Gamma_{dx}^{\mathrm{meas}}$} & \vline $< 71 \times 10^{-15}\,$m\,s$^{-2}$  & \\ \hline
{\bf Total of systematics in $\delta$ } & \vline $< 9 \times 10^{-15}$ & \\
\end{tabular}
\end{ruledtabular}
\end{table*}

Thermal effects are currently the dominant contribution to the systematic error. These were evaluated in a specific session where thermistors applied temperature variations at $f_{\rm EP}$ either to the electronic interface ($\Delta T_{\rm FEUU}$) or to the SU baseplate ($\Delta T_{\rm SU}$). The effect of these variations (or their gradients) on the differential acceleration signal is $\Gamma_{dx}^{\mathrm{meas}}(\mathrm{therm.})=(7\times10^{-11}\,m\,s^{-2}\,K^{-1}) \Delta{T_{\rm FEEU}}+(4.3\times10^{-9}\,m\,s^{-2}\,K^{-1}) \Delta{T_{\rm SU}}$.
The SU temperature sensitivity was more than 2 orders of magnitude larger than expected and far too large to be due to the radiometer effect or radiation pressure \cite{carbone07b} and thus must come from another source. Fortunately the maximum observed FEEU and SU temperature variations during 120 orbits were less than respectively $20 \times 10^{-6}$K and $15 \times 10^{-6}$K, about 2 orders of magnitude smaller than expected. The mean variation in fact was limited by the resolution of the probes, leading to the upper limit on the thermal systematic included in Table~\ref{tab:calibration}. Additional data could lower this upper limit.

The self gravity and magnetic effects have been evaluated by finite element calculation and found negligible compared to the previous error sources.

Fig. 2 shows the measurement spectrum for SUEP and SUREF. As expected, the measured noise varies as $f^2$ at high frequency; at low frequency, it varies as the $f^{-1/2}$ law expected for the damping noise of the gold wire. At $f_{\mathrm{EP}}$ the noise of the differential acceleration is dominated by this damping noise. It amounts to $5.6\times 10^{-11}\,$m\,s$^{-2}\,$Hz$^{-1/2}$ for SUEP and to $1.8\times 10^{-11}$\,m\,s$^{-2}\,$Hz$^{-1/2}$ for SUREF.

\begin{figure*}
\includegraphics[width=3.3in]{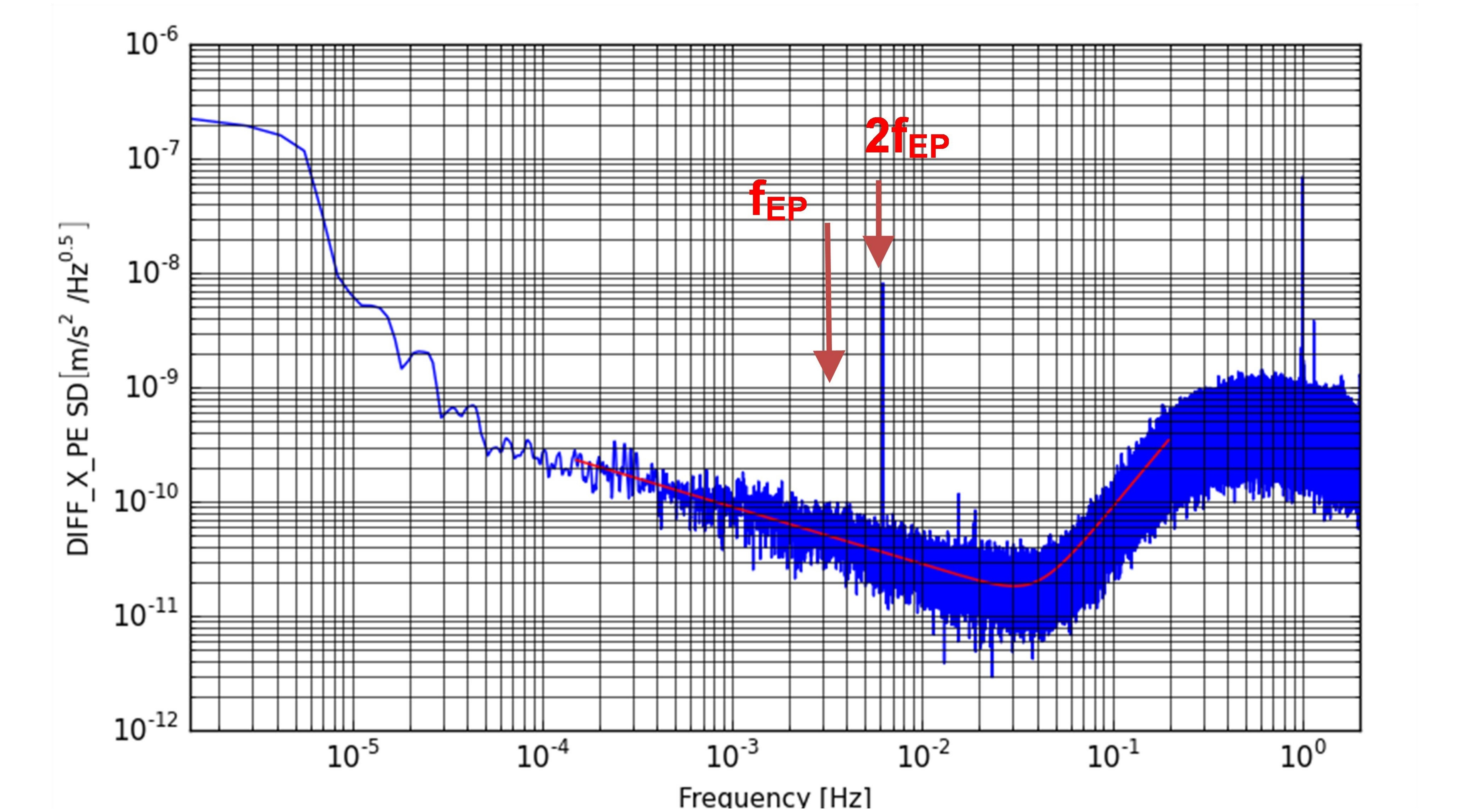}
\includegraphics[width=3.3in]{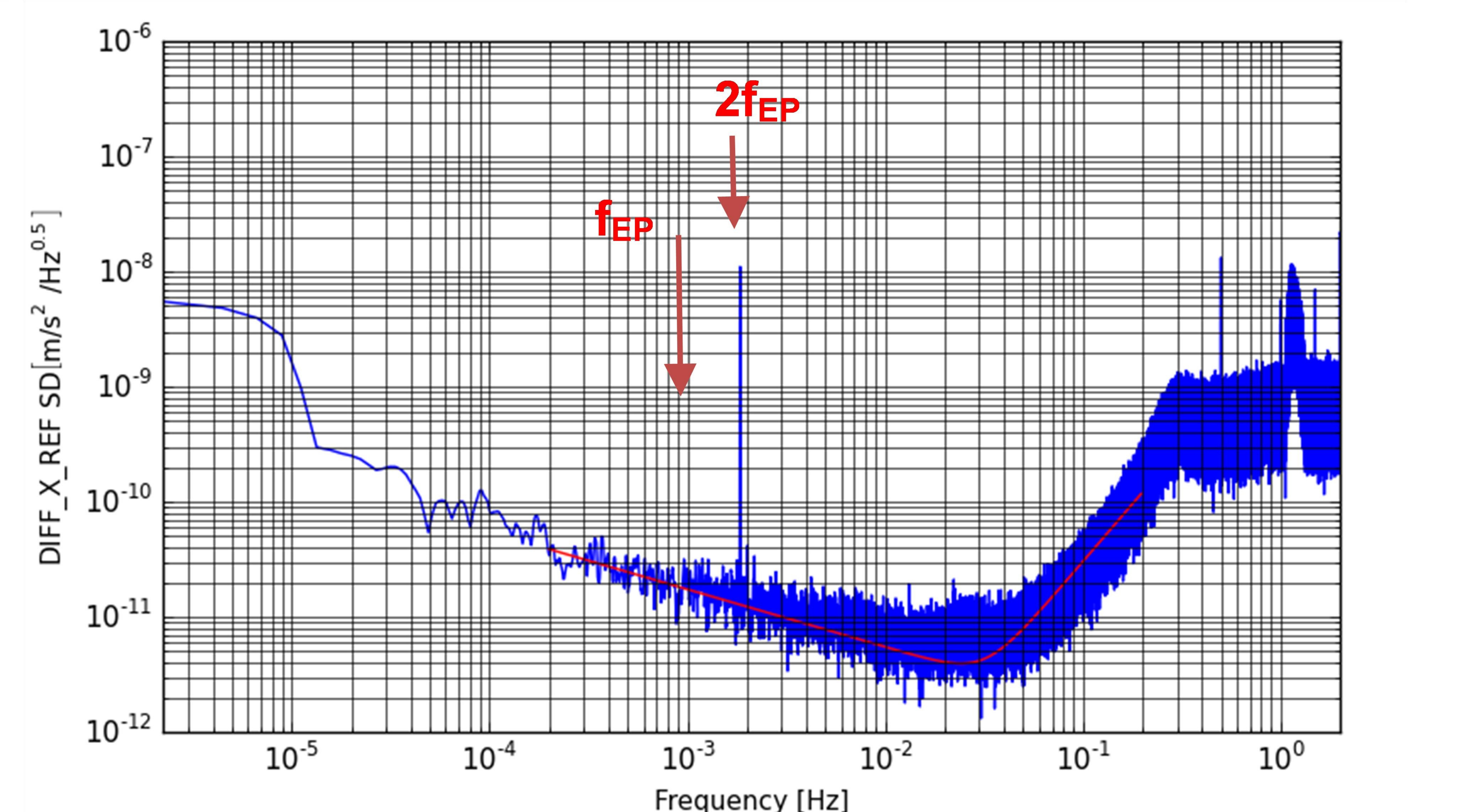}
\caption{\label{fig:psd}Square root of the measured PSD of the differential acceleration along X during the scientific session 218 with SUEP (left) and during the scientific session 176 with SUREF (right);
on left, $f_{\mathrm{EP}} = 3.1113 \times 10^{-3}$\,Hz,  $f_{\mathrm{orb}} = 1.6818\times 10^{-4}$\,Hz and satellite spin $= 2.9432\times 10^{-3}$\,Hz;  on right, $f_{\mathrm{EP}} = 0.9250\times 10^{-3}$\,Hz and satellite spin $= 0.5886\times 10^{-3}$\,Hz; the gravity gradient effect are clearly observed at $2f_{\mathrm{EP}}$. The red line is a power law fit to the spectrum. 
}
\end{figure*}

In the data used for this letter, the total amplitude of the differential acceleration FFT appears dominated by statistical signals over integration times lower than 62 to 120 orbits, respectively for SUREF and SUEP, as shown in Fig. 3: the blue line shows the evolution of the FFT amplitude at $f_{\mathrm{EP}}$ as the integration time (i.e number of orbits $N$) increases; the red line shows a $N^{-1/2}$ fit. The total FFT amplitude evolution appears inversely proportional to the square root of the integration time. A steady systematic effect would break this inverse proportionality law; for example a steady systematic effect (including a potential EP signal in SUEP) would show up as a constant offset. The results from both SUEP and SUREF are reaching sensitivities close to where no time dependent systematic effects should become apparent if they are present (without counterbalancing signal in SUEP) at the upper limit to the predictions shown in Table~\ref{tab:calibration}. 

\begin{figure*}
\includegraphics[width=3.3in]{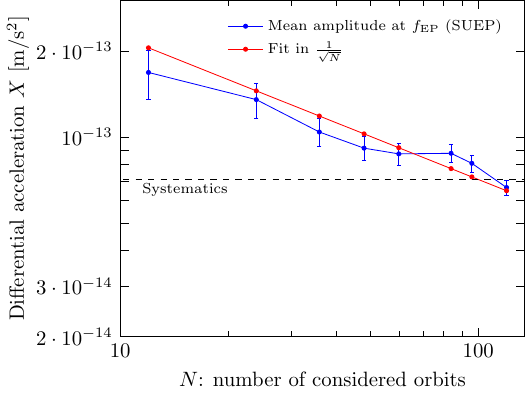}
\includegraphics[width=3.3in]{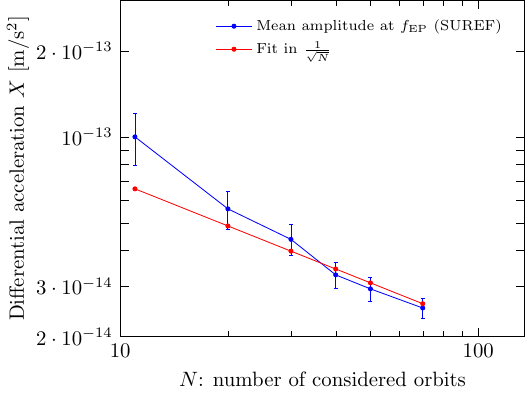}
\caption{\label{fig:time}Evolution of the mean amplitude of the FFT of the differential signal along X at $f_{\mathrm{EP}}$ as a function of integrating times (on left, from 12 to 120 orbits for the session 218 with SUEP and on right, from 11 to 62 orbits for SUREF). The mean of the FFT is computed as the average of the Fourier amplitudes over a narrow band of $10^{-4}$\,Hz around $f_{\mathrm{EP}}$. For SUEP, the black dashed line shows the estimated upper bound of the systematic errors given by the error assessment of Table~\ref{tab:calibration}.}
\end{figure*}

\section{\label{sec4}E\"otv\"os parameter estimation}

We simultaneously estimate the E\"otv\"os parameter $\delta(2,1)$ and the $\Delta_x$ and $\Delta_z$ miscenterings with a least-square fit based on Eq. (1) in the frequency domain. More precisely, $N$ equations (one per data point) in the time domain are converted into $N$ equivalent equations in the frequency domain through a Fourier transform; then the equation system is lightened by selecting the bands where the signal is expected (centered on $f_{\rm EP}$ for $\delta(2,1)$ and $2f_{\rm EP}$ for $\Delta_{x,z}$, with a $4\times  10^{-5}$\,Hz width~\cite{hardy13b}). 

The $1\sigma$ statistical errors are given by the $1\sigma$ uncertainty on the least-square estimate. The SUEP systematic error $9 \times 10^{-15}$ is given by the upper limit evaluation performed in Table~\ref{tab:calibration}.

The E\"otv\"os parameter for the SUEP instrument is obtained with 120 orbits (713,518\,s):
\begin{equation}
\delta \rm{(Ti,Pt)} = [-1 \pm 9 (\mathrm{stat}) \pm 9 (\mathrm{syst})] \times 10^{-15} \rm{at} ~1\sigma,
\end{equation}
with a goodness-of-fit $\chi^2_{\rm red}=1.17$

The test performed with the SUREF instrument over 62 useful orbits (368,650\,s) yields:
\begin{equation}
\delta \rm{(Pt,Pt)} = [+4 \pm 4 (\mathrm{stat})]\times 10^{-15} \rm{at} ~1\sigma,
\end{equation}
with $\chi^2_{\rm red} = 1.24$. 
This estimation is fully compatible with a null result (which is expected for this instrument), suggesting no evidence of systematic errors at the order of magnitude of $4\times 10^{-15}$ consistent with the SUEP conservative evaluation of Table~\ref{tab:calibration}.
To complete this analysis on the SUREF, specific sensitivity sessions are scheduled before the end of the mission in particular to detail the systematics.

\section{\label{sec5}Conclusion}

We have presented the first results on MICROSCOPE{'}s test of the Weak Equivalence Principle with conservative upper limits for some errors. Nevertheless this result constitutes an improvement of one order of magnitude over the present ground experiments~\cite{wagner12}. 
Forthcoming sessions dedicated to complete the detailed exploration of systematic errors will allow us to improve the experiment's accuracy.
Thousands of orbits of scientific measurements should be available by the end of the mission in 2018. The integration over longer periods of the differential accelerometer signal should lead to a better precision on the WEP test. MICROSCOPE will certainly take a step forward in accuracy, closer to the mission objective of $10^{-15}$ and bring new constraints to alternative gravity theories.

\begin{acknowledgments}
The authors express their gratitude to the different services involved in the mission partners and, in particular, CNES in charge of the satellite. They also thank Sandrine Pires (CEA Saclay) for numerous discussions and her help in the development of the data analysis pipeline. This work is based on observations made with the T-SAGE instrument, installed on the CNES-ESA-ONERA-CNRS-OCA-DLR-ZARM Microscope mission.
ONERA authors' work is financially supported by CNES and internal funding. Authors from OCA have been supported by OCA, CNRS and CNES.
ZARM authors'€™ work is supported by the German Space Agency of DLR with funds of the BMWi (FKZ 50 OY 1305) and by the Deutsche Forschungsgemeinschaft DFG (LA 905/12-1). \\
The authors would like to thank the referees for their useful comments.

\end{acknowledgments}

\bibliography{microscope3}
\bibliographystyle{unsrt}

\end{document}